\documentclass[11pt]{article}
\usepackage{fullpage}
\usepackage{graphicx}    % Support the \includegraphics command and options
\usepackage{amsmath}
\usepackage{esint}
\usepackage{array}       % For better arrays (eg matrices) in maths
\usepackage{verbatim}    % Adds better verbatim
\usepackage{color}
\usepackage{colortbl}
\usepackage{listings}
\usepackage{epstopdf}
\usepackage{subfigure}
\usepackage{url}
\usepackage{wrapfig}
\begin{document}
\title{From MPI to MPI+OpenACC:  Conversion of a legacy FORTRAN PCG solver for the spherical Laplace equation}
\author{Ronald M. Caplan\footnote{{\tt caplanr@predsci.com}}, Zoran Miki{\'c}, and Jon A. Linker \\
Predictive Science Inc., 9990 Mesa Rim Road Suite 170, San Diego, CA  92121}

\maketitle

\begin{abstract}
A real-world example of adding OpenACC to a legacy MPI FORTRAN Preconditioned Conjugate Gradient code is described, and timing results for multi-node multi-GPU runs are shown.  
The code is used to obtain three-dimensional spherical solutions to the Laplace equation.  Its application is finding potential field solutions of the solar corona, a useful tool in space weather modeling.
We highlight key tips, strategies, and challenges faced when adding OpenACC.
Performance results are shown for running the code with MPI-only on multiple CPUs, and with MPI+OpenACC on multiple GPUs and CPUs.
\end{abstract}

\section{Introduction}
\label{sec:intro}
The advent of hardware accelerators has had a significant impact on high-performance computing.  Performance normally only available on large super computers can be achieved in-house with a modest number of accelerator cards (sometimes at a much lower cost).  Accelerators also exhibit drastic improvements in power efficiency that allows the expansion of the capabilities of the largest high performance computing (HPC) systems.  In some cases, accelerators provide necessary performance that cannot be achieved with standard CPUs.
     
For software to take advantage of accelerators (or any multi-core system), they must be parallelized. Parallel programming is hard and parallel programming for accelerators is no exception. Originally, the only way to utilize accelerated computing was to write new code using APIs like CUDA \cite{cook2012cuda} and OpenCL \cite{kaeli2015heterogeneous}.  However, for legacy applications, such rewriting can be far too difficult in terms of time, effort, and education of the user base.  Additionally, because portability and longevity is of critical concern for large scientific codes, the adoption of accelerated programming to such applications has been somewhat slow.
   
Programming for accelerators through the use of directives has the potential to address many of the above concerns.  Since the directives are expressed as comments, a single source code can be used for both accelerated and non-accelerated runs.  Ideally, the addition of the directives does not require changing the current code base (although in practice, some changes are often inevitable).
 
OpenACC \cite{farber2016parallel} started as a temporary branch of OpenMP \cite{chapman2008using} to allow the use of accelerator hardware through directives.  Since then, it has evolved to be its own independent directive API. Implementers have added the ability to compile OpenACC code to multiple device types (including multiple vendors of GPUs and multi-core CPUs).  Its clean and simplified approach to directives has made it a desirable choice for adding accelerator capability to legacy codes.  We note that within the last few years, OpenMP has also added support for accelerators, but we do not explore that option here.
 
In this paper, we describe adding OpenACC to a legacy scientific code.  The application is a FORTRAN MPI code (POT3D) that solves the spherical Laplace equation.  It is used to compute potential field (PF) solutions of the Sun's coronal magnetic field given observations of the surface field.  Such solutions are very useful in solar physics and have space weather applications \cite{1970Natur226251S,wang1995solar,arge2003improved,toth2011obtaining}.  Our focus will be on using OpenACC to run the code on multiple accelerator devices (specifically, NVIDIA GPUs).
 
Our motivations/goals for transitioning POT3D from MPI to MPI+OpenACC are:
\begin{enumerate}
\item To efficiently compute solutions in-house using a modest number of GPUs that would normally require the use of a multi-CPU HPC cluster
\item To compute large solutions more quickly that current HPC resources allow
\item To use less HPC allocation resources (i.e. do an equivalent solve on fewer GPU node-hours than CPU node-hours)
\end{enumerate}
Although some uses of PF solutions do not necessitate the above improvements, another major motivation is that our 3D thermodynamic magnetohydrodynamic code (MAS \cite{MAS,downs2013probing}) has most of its run-time in equivalent PCG solvers as those found in POT3D.   The MAS code is used for very large simulations of the solar corona\footnote{e.g. see \url{www.predsci.com/eclipse2017} where MAS was used to predict the structure of the corona during the Aug. 21st, 2017 total solar eclipse} and heliosphere, and would greatly benefit from the above goals if they are found successful in POT3D.  Thus, adding OpenACC to POT3D can be viewed as a stepping-stone to potentially adding it to MAS.
   
Due to the legacy nature of the POT3D code, as well as the fact that not all researchers using the code have experience with GPU computing, a major objective is to apply OpenACC to POT3D with as few changes to the code as possible, and in a fully portable manner so that the same source code can be compiled and used on CPUs as before. 

The outline of the paper is as follows:  Sec.~\ref{sec:pot3d} describes the original POT3D code.  Then, in Sec.~\ref{sec:mpi2mpiopenacc}, we describe (using code examples) important tips, challenges, and solutions in adding OpenACC to POT3D.  The performance results using MPI and MPI+OpenACC on CPU and GPU multi-node clusters is shown in Sec.~\ref{sec:results}.  Performance portability is discussed in Sec.~\ref{sec:portable} and we conclude in Sec.~\ref{sec:conclude}.

\section{POT3D}
\label{sec:pot3d}
POT3D is a FORTRAN code that computes potential field solutions to approximate the solar coronal magnetic field using observed photospheric magnetic fields as a boundary condition.  It has been (and continues to be) used for numerous studies of coronal structure and dynamics \cite{linker2016empirically,titov20122010}.

\subsection{Potential Field Model}
A PF is a force-free and current-free magnetic field.  Setting the current in Maxwell's equation ($\nabla \times \vec B =\mu_0\,\vec J$) to zero ($\vec J=0$) leads to solutions of the form $\vec B = \nabla \Phi$, where $\Phi$ is the scalar potential.  Combined with the divergence-free condition ($\nabla\cdot\vec B = 0$), this yields a Laplace equation for $\Phi$:
\begin{equation}
\label{eq:laplace}
\nabla^2\Phi=0.
\end{equation}
The lower boundary condition at $r=R_\odot$ (the solar surface) is set based on an observation-derived surface-map of $B_r$ as
\begin{equation}
\label{eq:laplace_bcr0}
\left.\frac{\partial \Phi}{\partial r}\right|_{R_{\odot}} = \left.B_r\right|_{R_\odot}.
\end{equation}
Depending on the needs of the application, the upper boundary ($r=R_1$) is set to either a `closed wall' ($\left.B_r\right|_{R_1}=0$) or `source surface' ($\left.\Phi\right|_{R_1}=0$) condition. The $\phi$ direction uses a periodic boundary ($\left.\Phi\right|_{\phi=0}=\left.\Phi\right|_{\phi=2\,\pi}$), while polar boundary conditions are set using the average
\begin{equation}
\label{eq:laplace_bcpole}
\left.\Phi\right|_{\theta_{0/\pi}}=\frac{1}{2\,\pi}\int_{\phi=0}^{\phi=2\,\pi}\left.\Phi\right|_{\theta_{0/\pi}\pm\epsilon}\,d\phi,
\end{equation}
where $\epsilon$ is set to half the cell width next to the pole.  Once Eq.~(\ref{eq:laplace}) is solved, we set $\vec B = \nabla \Phi$ to get the magnetic field result.

\subsection{Finite-Difference}
POT3D uses a globally second-order finite-difference method on a nonuniform logically-rectangular spherical grid \cite{NUG_PLAY_1992}. In this scheme, Eq.~(\ref{eq:laplace}) takes the form 
\begin{alignat}{2}
\label{eq_diffuse_desc1}
\nabla^2 \Phi_{i,j,k} &\approx \frac{1}{\Delta r_i}\left[
\frac{\Phi_{i+1,j,k}-\Phi_{i,j,k}}{\Delta r_{i+\frac{1}{2}}}
-\frac{\Phi_{i,j,k}-\Phi_{i-1,j,k}}{\Delta r_{i-\frac{1}{2}}}
\right] \notag \\
&+ \frac{1}{\sin\theta_j\,\Delta\theta_j}\left[
 \sin\theta_{i,j+\frac{1}{2}}\,\frac{\Phi_{i,j+1,k}-\Phi_{i,j,  k}}{\Delta\theta_{j+\frac{1}{2}}}
-\sin\theta_{i,j-\frac{1}{2}}\,\frac{\Phi_{i,j,  k}-\Phi_{i,j-1,k}}{\Delta\theta_{j-\frac{1}{2}}}
\right] \notag \\
&+\frac{1}{\sin^2\theta_j\,\Delta\phi_k}\left[     
 \frac{\Phi_{i,j,k+1}-\Phi_{i,j,k  }}{\Delta\phi_{k+\frac{1}{2}}}
-\frac{\Phi_{i,j,k  }-\Phi_{i,j,k-1}}{\Delta\phi_{k-\frac{1}{2}}}
\right]=0, \notag
\end{alignat}
which can be represented in matrix form as 
\begin{equation}
\label{eq:pot3d}
{\bf A}\,\Phi = 0.
\end{equation}
The matrix ${\bf A}$ is stored in the DIA sparse format \cite{DIACSR} for the inner grid points, while the boundary conditions are implemented matrix-free.

\subsection{Preconditioned Conjugate Gradient}
The matrix ${\bf A}$ of Eq.~(\ref{eq:pot3d}) is large, symmetric, and sparse.  A common method for solving such systems is the Preconditioned Conjugate Gradient (PCG) method \cite{IterativeMethods_SAAD_Book}.  

Effective use of the PCG method requires selecting an inexpensive yet efficient preconditioner (PC) \cite{PC_iterative_survey}.  For POT3D, we use two communication-free PCs: 1) A point-Jacobi/diagonal-scaling (PC1) which uses the inverse of the diagonal of ${\bf A}$, and 2) A non-overlapping domain decomposition with zero-fill incomplete LU factorization (PC2) \cite{IterativeMethods_SAAD_Book}.  PC1 has a fully vectorizable and inexpensive formulation and application, but it is limited in its effectiveness at reducing iterations. PC2 is much more expensive to formulate and apply, but can be much more effective. In POT3D, we use both PC1 and PC2 since it is possible for PC1 to outperform PC2 (e.g. when solving problems requiring very few iterations, when using hardware that is very efficient for vectorizable algorithms, or when the ILU0 suffers `breakdown' \cite{ILU_breakdown}). 

The PCG iterative solver consists of the following main operation types:
\begin{itemize}
\item Vector-vector operations (e.g. ${\vec z} = a\,{\vec x}+b\,{\vec y}$)
\item Inner products (e.g. $a={\vec x}\cdot {\vec y}$)
\item Matrix-vector products (e.g. ${\vec y}={\bf A}\,\vec x$)
\item Application of the preconditioner
\end{itemize}
The vector-vector operations are fully vectorizable and are expected to run very well on GPUs.  The inner products are vectorizable in their local sums, but can create poor parallel scaling between subdomians due to their collective/synchronous nature.  The matrix-vector products are also fairly vectorizable. The application of the preconditioner for PC1 is a simple vector operation.  For PC2, the LU matrix is stored in a memory-optimized CSR format \cite{CSRopt} for better access patterns, but the triangular solves are done with a standard algorithm \cite{IterativeMethods_SAAD_Book} which is not vectorizable.  For more details on the implementation of PCG in POT3D, see Ref.~\cite{ASTRONUM16}.

\subsection{MPI Parallelism}
\label{sec:mpi}
POT3D is parallelized using MPI in a domain decomposition manner.  Each MPI rank takes one subsection of the grid (as cubed as possible) and treats it as its own local domain for all operations.  The only MPI communication needed are point-to-point messages for the local boundaries in the matrix-vector product, and collective operations for the inner products and polar boundary conditions.  Initial and final collectives are also used to decompose the domain and collect the solution for output.  

Some MPI scaling issues of the code include the polar boundary conditions of Eq.~(\ref{eq:laplace_bcpole}) creating load imbalance due to the implicit synchronization of the polar subdomains, and the global synchronization of the inner products not allowing small load imbalances of the subdomian sizes and/or hardware performance variations to overlap/equalize.

To allow the MPI library to optimize the point-to-point communication as best as possible, the MPI ranks are reordered using a Cartesian topology which mirrors the stencil pattern of the decomposition including the periodicity in the $\phi$-direction.

\section{MPI to MPI+OpenACC}
\label{sec:mpi2mpiopenacc}
In this section we describe some key steps in adding OpenACC to POT3D.  There are many thorough tutorials and examples for implementing OpenACC including the OpenACC website\footnote{\url{www.openacc.org}} and Ref.~\cite{farber2016parallel}.  The scope of this paper is to only focus on a few of the more important and/or not-so-common strategies and challenges that we encountered.

\subsection{Avoiding GPU-CPU data movement}
\label{sec:dataregion}
One of the biggest challenges when using accelerator hardware with its own localized memory (such as GPUs) is efficiently handling data movement.  Transferring data between the CPU and GPU is very slow, and can degrade performance to the point of the code becoming unusable.  The recent development of unified memory, which allows the run-time environment to efficiently handle the data movement automatically, has potential to alleviate this concern (and also allow for running problems that are larger than the accelerator device's memory).  However, since manually controlling the data movement often yields improved performance, we do not use unified memory in POT3D.    

The most efficient way to handle data transfers is to keep the data on the GPU throughout the entire computation as much as possible.  In OpenACC this can be accomplished by the use of unstructured data regions.  With a simple pragma, one can transfer the data to/from the GPU and utilize it in OpenACC compute constructs using the {\tt present} clause.  When the data is needed on the CPU (e.g. for IO), a simple {\tt update} clause can transfer the data back.  The following pseudo code illustrates the use of unstructured data regions:
\begin{lstlisting}
    @Initialize $\Phi$ on CPU@
!$acc enter data copyin(phi)
    @Solve system on GPU using device copy of $\Phi$@
    @with the present clause@
!$acc update self(phi)
    @The result in $\Phi$ is now available to the CPU for IO@
!$acc exit data delete(phi)
    @$\Phi$ is now cleared out of GPU memory@
\end{lstlisting}

\subsection{The use of Kernels for FORTRAN array-syntax}
The prototypical compute region that OpenACC is designed for is a loop construct (such as a {\tt do} loop in FORTRAN or a {\tt for} loop in C).  However, FORTRAN's built-in array syntax is often used in place of explicit {\tt do} loops for its simplicity as well as allowing the compiler to vectorize/optimize easily.  POT3D contains multiple uses of array-syntax code and although one could expand these into explicit loops, doing so would require a lot of code rewriting.  

In order to use OpenACC with these array operations, one can use the {\tt kernels} pragma around the code in question as shown here:
\begin{lstlisting}
!$acc kernels present(x,p,ap,r)
        x=x+alphai*p
        r=r-alphai*ap
        ap=r
!$acc end kernels 
\end{lstlisting}

However, this technique has a few drawbacks.  For one, this gives full control to the compiler on how to order the underlying loops and attach them to CUDA block dimensions (i.e. deciding on the {\tt gang}, {\tt worker}, and {\tt vector} topology and sizes), that can lead to sub-optimal decompositions.  

Another issue is that most compilers will implement {\tt kernels} in the safest way possible, invoking a separate GPU kernel for each line of array operation code.  This can cause kernel launch overheads that ,in some cases, can non-trivially reduce performance.  In contrast, if one expands the code into explicit loops and uses the alternative {\tt parallel} pragma, all operations can be compiled into a single kernel.

Finally, we note an issue (independent of OpenACC) that to adhere to the Fortran standard, the right-hand-side of array-syntax operations must be able to be evaluated first, and then copied to the left-hand-side.  In most cases, the compiler will optimize this and directly perform the operation and storage.  However, if the arrays were declared with the {\tt pointer} attribute, the compiler must be alias-safe and create a temporary array to store the right-hand-side results and then do a copy.  The best way to avoid this is to either use {\tt allocatable, target} when declaring arrays, or use the {\tt contiguous} descriptor when declaring pointers that will be used with stride-1 data\footnote{See M. Wolfe PGI Insider Vol 6 (3) \url{http://www.pgroup.com/lit/articles/insider/v6n3a4.htm}}.  

\subsection{Multiple Devices} 
The use of multiple GPU devices can be quite challenging to implement, especially across a large cluster.  One typically needs to use multiple threads or processes to invoke each device.

Fortunately, POT3D's MPI domain decomposition was implemented in a modular fashion in that the MPI nature of the code is transparent to the developer except in the few subroutines where MPI calls are necessary.  Thus, each MPI rank appears to compute on its own sub-domain problem.  Because of this, adding multi-GPU capability in the simplest manner possible was surprisingly straight-forward and could be done with minimal changes to the code.

We added a new input parameter ({\tt ngpus\_per\_node}) to the code that indicates the number of GPUs per node, which is then used to set the MPI rank's device using the OpenACC pragma {\tt device\_num}:
\begin{lstlisting}  
      igpu=MODULO(iprocw,ngpus_per_node)
!$acc set device_num(igpu)  
\end{lstlisting}
All code after the above pragma uses a unique GPU.  It is necessary in this case that there is one MPI rank per GPU on a node (or, one could oversubscribe each GPU, but this is more complicated because it requires special driver settings that may or may not be activated on a particular system).  It is important to note that the above code assumes linear affinity of the MPI ranks across the cluster.  Since we allowed the ranks to be rearranged due to the Cartesian topology (see Sec.~\ref{sec:mpi}), we had to make sure to use the rank ID from the original {\tt MPI\_COMM\_WORLD} communicator.  The assumption of linear affinity could cause portability problems (as HPC systems and libraries could have other affinities by default).  A more portable way to achieve the above is to use the shared communicators introduced in MPI-3\footnote{\url{http://mpi-forum.org/mpi-31/}}.  These communicators are defined on each node and allows the code to know how many ranks there are in the node, and numbers them linearly.  We do not use shared communicators in POT3D out of concern for portability, as the MPI libraries on some of the HPC systems we use are not updated to MPI-3.  

The use of multiple devices across the cluster requires that the MPI calls for point-to-point and collective operations transfer data from one GPU to another GPU (which could reside on another node in the cluster).  To achieve this without changing the code, we utilize CUDA-aware MPI.

CUDA-aware MPI libraries detect that a pointer given to an MPI call is a GPU device pointer and sets up the GPU memory transfers automatically.  How this is done depends on the MPI library's implementation and the available hardware features (e.g. GPU-Direct, NVLink, etc.).
In OpenACC, the {\tt host\_data use\_device} pragma is used to tell the compiler to use the device version of the pointer in the succeeding subroutine calls (assuming the data has already been transfered to the GPU).  For example, 
\begin{lstlisting}
!$acc host_data use_device(a0)
    call MPI_Allreduce (MPI_IN_PLACE,a0,n,ntype_real,
   &                    MPI_SUM,comm_phi,ierr)
!$acc end host_data
\end{lstlisting}

Although simple to code, the use of CUDA-aware MPI relies heavily on the HPC system and MPI libraries to be up-to-date and set up properly.  We experienced difficulties on two separate HPC systems when trying to use CUDA-aware MPI. In one case the compiler/libraries simply needed an update, while in the other case, an undetected bug was eventually discovered in the CUDA-aware implementation of the MPI library.  How the MPI library implements CUDA-aware MPI is also an issue since it may do so in a non-optimized way that the programmer has no control over.

\subsection{Calling the cuSparse library from FORTRAN}
As mentioned in Sec.~\ref{sec:pot3d}, the application of the PC2 preconditioner as implemented in POT3D is not vectorizable.  This severely limits the ability to efficiency compute it on a GPU using OpenACC pragmas (in fact, the resulting code would compute serially on a single GPU thread!).  Since running serially on the GPU or transferring the data back and forth to compute it on the CPU are both non-starters in terms of performance, we link the code to an external library.  

Although linking to libraries is often an excellent idea when needing algorithms that have already been optimized by experts, in our case this is undesirable because it breaks the portability of the code by requiring numerous code changes and additions including multiple build options.  One could mitigate the problem with the use of {\tt \#ifdef}s, but for legacy codes, changing the look, feel, and installation methods can cause users to opt out of using the new code.  For the purposes of this study, we made two branches of POT3D:  One branch only supports OpenACC acceleration when using PC1, while the other uses external libraries to support acceleration when using PC2 as well.  The first branch remains fully portable, while the second loses portability.

The application of PC2 requires solving two successive triangular sparse matrix equations.  Solving triangular matrix equations in a vectorized manner is very difficult. The cuSparse library provided by NVIDIA implements two such algorithms\footnote{See the {\tt csrsv} and {\tt csrsv2} functions described at \url{ http://docs.nvidia.com/cuda/cusparse/index.html#cusparse-level-2-function-reference}}, each having their own options and parameters.  Although we implemented calls to both algorithms, considering that in our case we did not find noticeable differences in performance between the two, we only use the first algorithm.  

The cuSparse library is written in C, and to allow FORTRAN users to use the library directly, FORTRAN wrappers are provided.  However, in order to minimize the amount of added code to original POT3D FORTRAN source code as much as possible, we did not use these wrappers, but instead wrote a small C function that calls the cuSparse library.  We then call this function from the FORTRAN code, which requires very few lines of added code.  Calling C functions from FORTRAN used to be somewhat difficult as each compiler had its own requirements and syntax.  Fortunately, a fully portable method of calling C functions is now part of FORTRAN as of the 2008 standard.  A full example can be found on pg. 283 in Ref.~\cite{farber2016parallel}, but for demonstration purposes, we reproduce some of the code here.  The C code we wrote contains the following function which calls the cuSparse LU solvers:
\begin{lstlisting}
void lusol_cusparse(double* x,double* CSR_LU,
                   int* CSR_I,int* CSR_J,int N,int M){
  cusparseDcsrsv_solve(cusparseHandle,
       CUSPARSE_OPERATION_NON_TRANSPOSE,N,&one,
       L_described,CSR_LU,CSR_I,CSR_J,L_analyzed,x,x);
  cusparseDcsrsv_solve(cusparseHandle,
       CUSPARSE_OPERATION_NON_TRANSPOSE,N,&one,
       U_described,CSR_LU,CSR_I,CSR_J,U_analyzed,x,x);
  cudaDeviceSynchronize();
}
\end{lstlisting} 
Within the FORTRAN code, an interface is created declaring the FORTRAN subroutine of the same name to be linked to the C code:
\begin{lstlisting}
interface
 subroutine lusol_cusparse(x,CSR_LU,CSR_AI,CSR_AJ,N,M)
&  BIND(C, name="lusol_cusparse")
   use, intrinsic :: iso_c_binding
   integer(C_INT), value :: N,M
   type(C_PTR), value :: x,CSR_LU,CSR_AI,CSR_AJ
 end subroutine lusol_cusparse 
end interface
\end{lstlisting}
Within the code, this routine can now be called (with variables cast as C-pointers and C-scalars) as follows:
\begin{lstlisting} 
!$acc host_data use_device(x,a_csr,a_csr_ja,a_csr_ia)
      call lusol_cusparse(C_LOC(x(1)),C_LOC(a_csr(1)),
     &                    C_LOC(a_csr_ia(1)),
     &                    C_LOC(a_csr_ja(1)),cN,cM),cM)
!$acc end host_data
\end{lstlisting} 
where we have once again used the {\tt host\_data} pragma of OpenACC to indicate that the device pointers should be used in the function call.

The above code samples are for applying the PC2 preconditioner.  To use the cuSparse calls, we also needed to set up initializations of cuSparse, using global variables in the C code which are then available to successive calls.  To minimize code changes to POT3D, we initialize the ILU0 matrix on the CPU and transfer it to the GPU for use in cuSparse.

To build the code, we use {\tt nvcc} to compile the C code which calls cuSparse, and then link the resulting object file to the compilation of POT3D.

\subsection{Additional considerations}
Here we describe a collection of smaller additional considerations, which were of less importance in our case, but may be more relevant to other codes.

The unstructured data regions described in Sec.~\ref{sec:dataregion} work very well for global arrays (or arrays in a module).  Often in codes, a temporary array is initialized (statically or dynamically) within a subroutine and only has a local scope.  Such arrays can be allocated on the GPU, but they then need to invoke a kernel loop to initialize them (or they can be initialized on the CPU, but then they would need to be transfered to the GPU).  In POT3D, in order to avoid having to allocate and initialize temporary arrays on the device over and over again, we chose to modify the code to make those few arrays global in scope.  

Another issue involving data movement is that occasionally, only a few (or one) values of an array are needed in a compute loop.  Although one can simply transfer the whole array to the device, this creates overhead (since the array could be accessed through cache rather than a register) and in some cases, the array might be quite large.  In POT3D, we had an example of a single value of an array ($\mbox{dr}(1)$) being used in a loop.   We chose to define a new scalar value ($\mbox{dr1}=\mbox{dr}(1)$) instead of transferring the array in the following kernel:
\begin{lstlisting}
!$acc kernels present(x,br0)
       x(1,2:ntm1,2:npm1)=x(2,2:ntm1,2:npm1)
      &            -vmask*br0(2:ntm1,2:npm1)*dr1
!$acc end kernels
\end{lstlisting}

Another consideration is that on nested loops, the choice of {\tt gang}, {\tt worker} and {\tt vector} topology and sizes can somewhat affect performance (note that the compiler is allowed to override the programmer's choices).  For POT3D, we chose what seemed reasonable for the GPUs we planned to run on (using examples from Ref.~\cite{farber2016parallel} and others).  However, we have found that simply allowing the compiler full freedom to choose the topology typically yields close to optimal performance (doing so is also beneficial for performance portability {--} see Ch. 7 of Ref.~\cite{OpenACCBook2}). 

Another issue affecting performance is the choice of compiler flags.  While this is obviously compiler dependent, it is important to keep in mind.  For example, we were able to obtain a sizable performance gain by simply invoking CUDA 8.0 instead of the default (7.5) in the PGI compiler (version 16.10).  

For codes that require frequent data transfers between GPU and CPU, asynchronous data movement and compute regions can help to fill the device pipeline by overlapping computation of different kernels and/or data movement.  In POT3D, we do not make much use of asynchronous clauses since the form of the PCG algorithm does not lend itself to communication-computation overlap, and we only transfer data between the GPU and CPU at the start and the end of the solve.

\subsection{Profiling}
\label{sec:profile}
\begin{figure}[htbp]
\centering
\includegraphics[width=\textwidth]{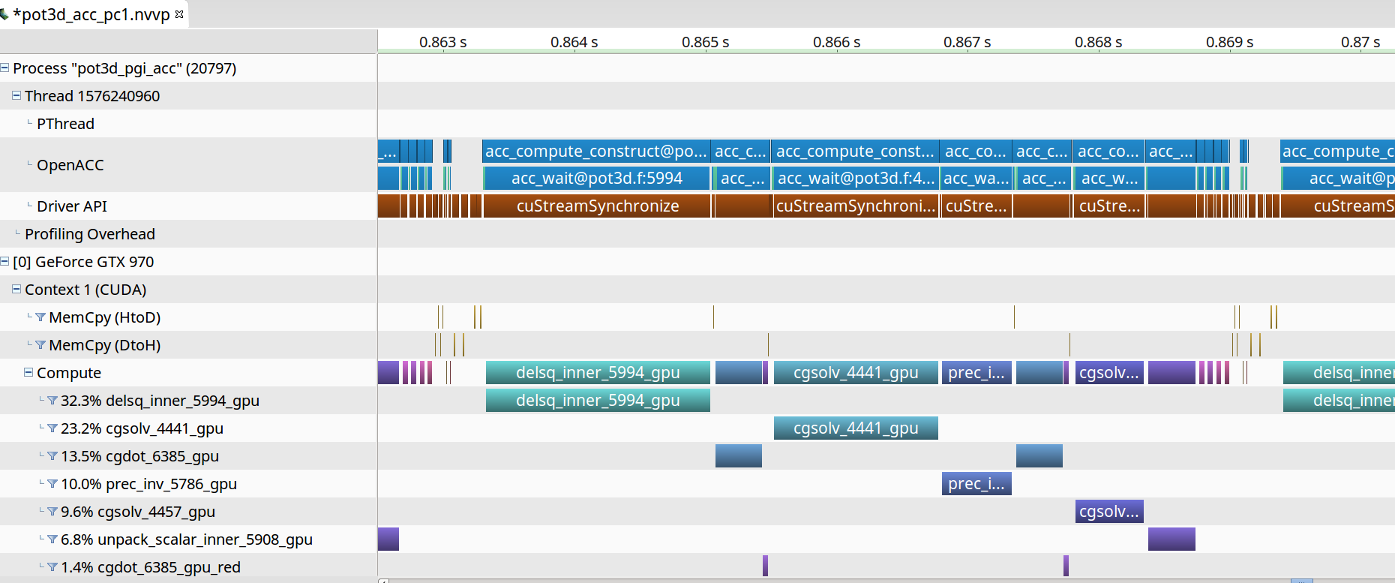}
\\
\vspace{0.5cm}
\includegraphics[width=\textwidth]{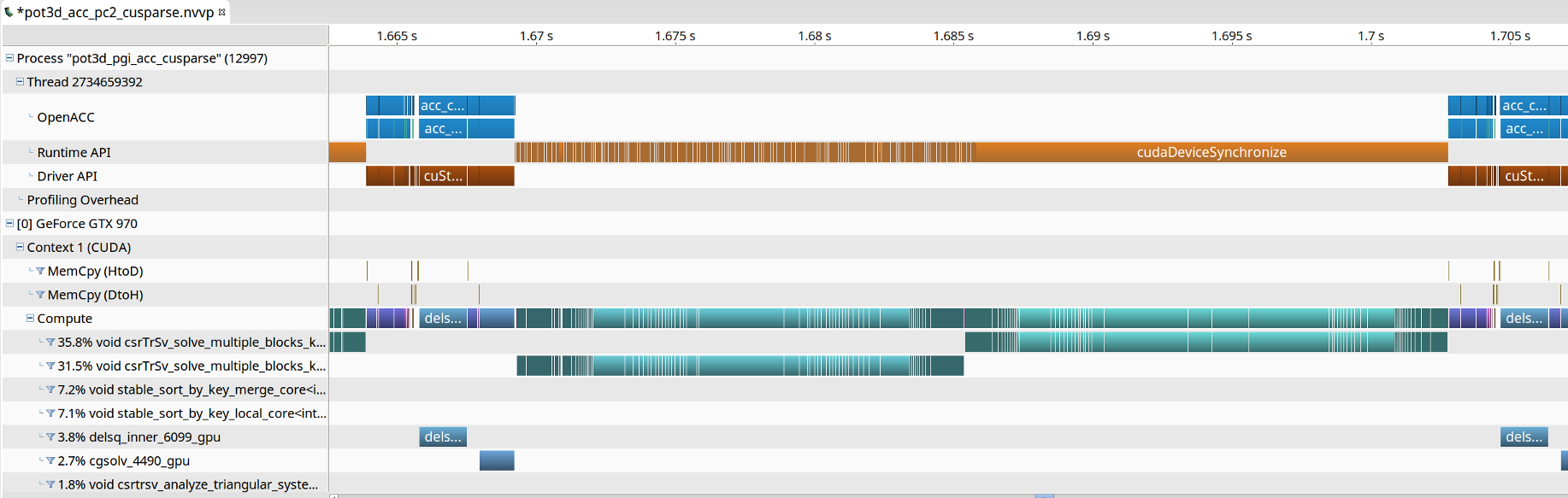}
\caption{Profiling results of one iteration of POT3D's PCG solver using OpenACC on a single NVIDIA GeForce 970 GPU for PC1 (top) and PC2 (bottom). \label{fig:profiling}} 
\end{figure}

To ensure good performance, it is often critical to perform profiling tests.  In the case of using OpenACC with GPUs, this is even more important as unwanted/unexpected GPU-CPU memory transfers can drastically reduce performance.  In Fig.~\ref{fig:profiling}, we show screenshots of profiling (using {\tt pgprof}) the OpenACC POT3D run on a single NVIDIA GeForce 970 GPU on a small test case ($120\times120\times240$) with both PC1 and PC2 preconditioners.  Our goal in this simple profiling is not meant to provide a detailed performance analysis, but rather to 
get an idea of how the code is running and if there are any glaring performance problems.    
We see that there are no unexpected GPU-CPU transfers and little overhead.  However, in the case of PC2, the cuSparse library's LU-solvers are much slower than the matrix-vector product.  While this is true in the CPU case as well, it is a relatively larger fraction here.   One reason why this might be the case is that the algorithms for LU-solves in cuSparse are most efficient for matrices with a large number of non-zeros per row.  In a case of a simple 7-point stencil, the solver does not seem to be able to efficiently extract enough parallelism.  It is possible that storing the matrices in ELL format \cite{CSRopt} may have yielded better results, but this would have required even more code changes to POT3D.

\subsection{Summary of effort and portability}
\label{sec:effort}
Adding OpenACC to POT3D for PC1 required minimal code changes and the addition of $67$ OpenACC pragmas equaling $230$ additional lines of code. Compared to POT3D's original line count ($7668$), this is only a $3\%$ modification.  The resulting code is fully portable in that it can be compiled in the original MPI-only manner for CPUs, or with OpenACC enabled for running on multiple GPU (or other) devices. 

For the PC2 preconditioner, the necessity of calling the cuSparse library increased the required amount of modifications to the code.  The OpenACC pragmas and additional FORTRAN statements added up to 335 lines of code, while the additional C code used to call cuSparse was another 181 lines.  Combined, this equates to a modification of $6.3\%$.  While still small, this is over twice the amount than was  required for PC1 alone.  More importantly, adding cuSparse broke portability by requiring the cuSparse library (and a new build script) to compile the code, and limiting the OpenACC implementation to NVIDIA GPUs.  This lack of portability could be overlooked if the PC2 performs substantially better than PC1 but, as we will show in the next section, this is not the case.

In the current development, we did not expend excessive time on detailed performance tuning. We wanted to see how much `time-to-solution' performance could be achieved using OpenACC with minimal development time (since reducing development time is one of the key proposed benefits of using OpenACC for GPU computing). 

\section{Performance Results}
\label{sec:results}
In this section we show some performance results for running POT3D with OpenACC\footnote{A MATLAB script containing all timing data shown in this section is available at \\ \url{www.predsci.com/~caplanr/pot3d}}.  We choose a problem size that is quite large so that we can test the scaling to multiple devices.  We emphasize that the results presented here are not necessarily indicative of results one can expect in general, but are rather offered as a single real-world example.  Variations in optimizations, algorithm choices, and hardware setups can vary performance (sometimes significantly).

We also emphasize that our goal is not to compare the performance of the different hardware architectures under identical software stacks, but rather to compare the time-to-solution performance when utilizing them.  Therefore, we use the compiler and compile options best suited (or best available) for each hardware setup.

\subsection{Test Case}
\begin{figure}[htbp]
\centering
\includegraphics[height=2.00in]{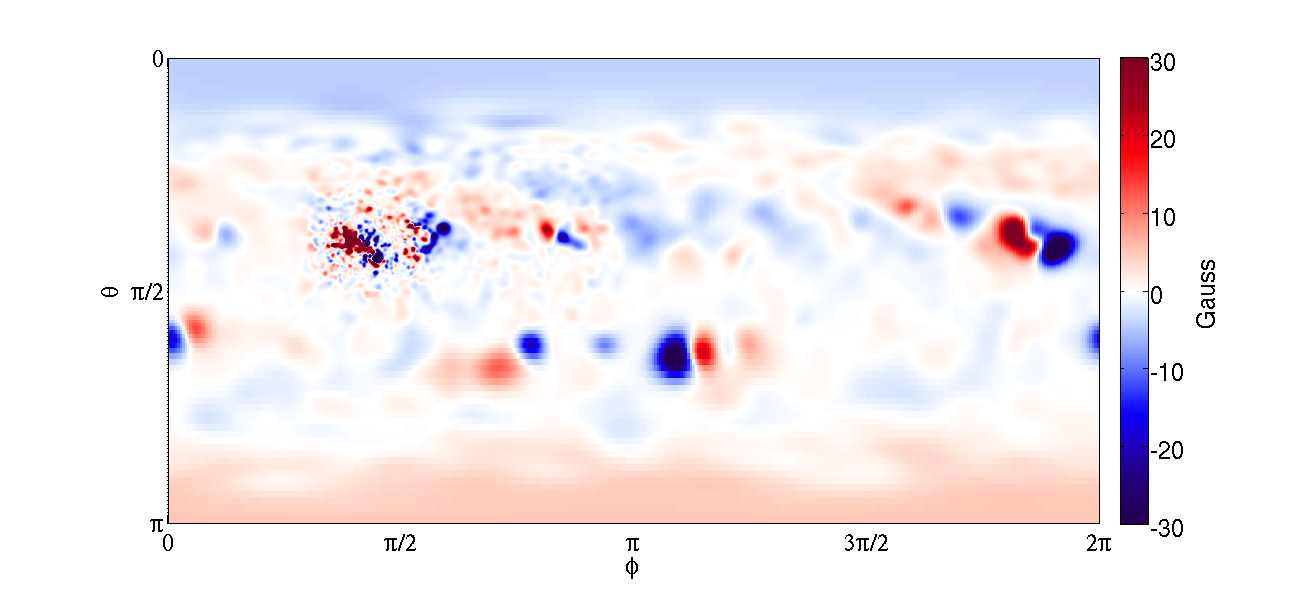}
\includegraphics[height=2.00in]{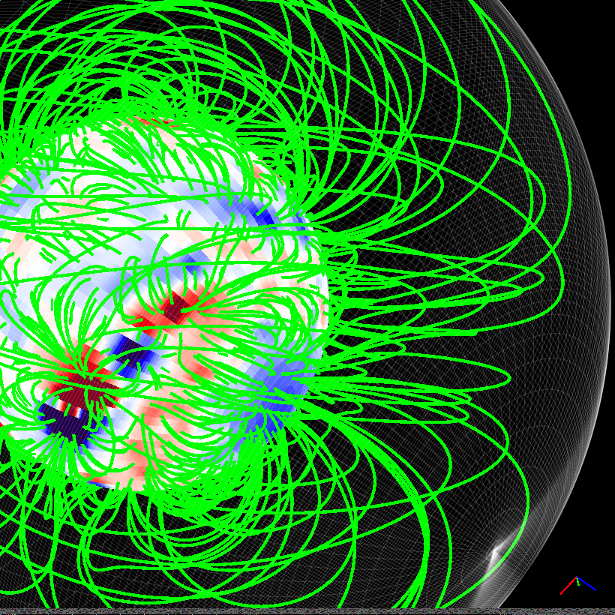}
\caption{Input surface magnetic field boundary condition (left) and selected field lines from the potential field solution (right) for the chosen test case.\label{fig:testcase}} 
\end{figure}

Our test case utilizes the solar surface magnetic field boundary described in Ref.~\cite{titov20122010}, and computes the PF solution with a mesh size of $(N_r\times N_\theta\times N_\phi)=(200\times 720\times 1440)$ yielding a total of over 200 million grid cells. Fig.~\ref{fig:testcase} shows the boundary magnetic field and the resulting potential field solution.  
The solver is computed in double precision and iterated to a relative tolerance of $10^{-9}$ and takes $19995$ and $2563\rightarrow 3221$ iterations when using PC1 and PC2 respectively (the range of iterations using PC2 is because it becomes less effective as the number of processors increases due to its non-overlapping implementation).  For all timings, we subtract off the time spent in IO, but do include the start-up time (e.g. the ILU0 decomposition and initial CPU{-}GPU data transfers).  This start-up time is typically negligible in light of the number of iterations in the solve.  

For a baseline reference, we ran the test serially on a single CPU core.  To maintain the practical `time-to-solution' approach, we do not use a single core on the HPC CPU nodes we will use, as they are often slower per core then a typical desktop CPU.  We compiled the code using GNU 5.4 with `-O3' and ran it using the PC1 algorithm on a single core of an Intel Xeon E5-1650v4 3.6GHz desktop CPU yielding a run-time of $61576$ seconds (over 17 hours).  

\subsection{HPC Resources}
To analyze the original CPU MPI-only scaling of POT3D (and using MPI+OpenACC compiled to x86 multi-core) we use the Comet supercomputer at the San Diego Supercomputing Center.  Comet consists of 24-core Xeon E5-2680v3 2.5GHz Haswell (two 12-core sockets) compute nodes linked by Infiniband\footnote{For full specifications, see \url{www.sdsc.edu/support/user_guides/comet.html}}.  The maximum allowed number of CPUs to use in one run (for standard allocations) is 72 nodes (1728 cores).  For the MPI CPU runs, we use the Intel 2015.2.164 compiler with the MVAPICH 2.1 MPI library, while for the MPI+OpenACC x86 multi-core runs, we use the PGI 17.5 compiler with the same version of the MVAPICH library.

For testing MPI+OpenACC runs on NVIDIA GPUs we used the PSG cluster at NVIDIA which has 4 Pascal P100 GPUs per node.  The code was compiled with the PGI compiler (version 16.10 + CUDA 8.0) and the OpenMPI 1.10.2 MPI library.  

\subsection{CPU performance with MPI and MPI+OpenACC}
\begin{wrapfigure}{r}{0.5\textwidth}
\centering
\includegraphics[width=0.5\textwidth]{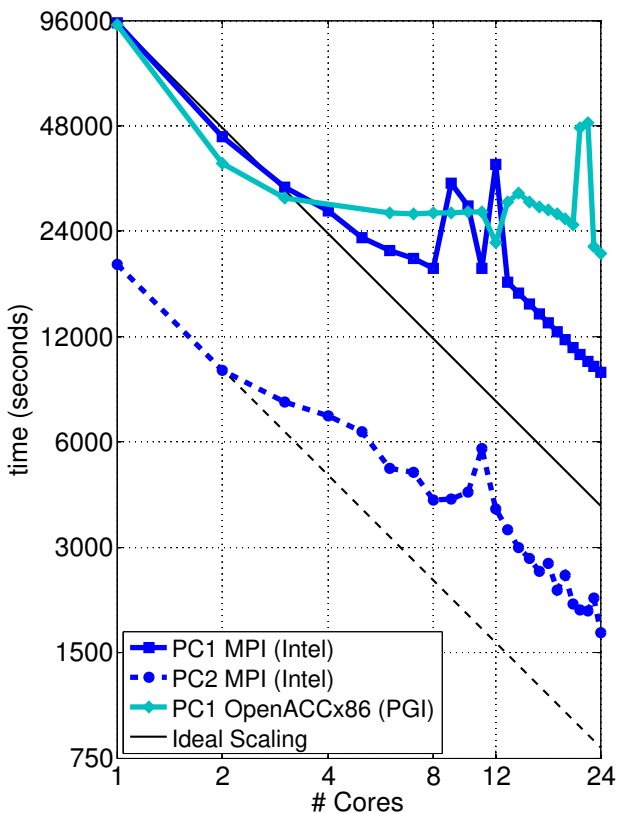}
\caption{CPU timing results of running POT3D on the test case with MPI-only and OpenACC compiled to x86 multicore on a single CPU node.\label{fig:time_cpu}} 
\end{wrapfigure}

The performance results for running POT3D on a single CPU node are shown in Fig.~\ref{fig:time_cpu}.  The scaling of the code is not ideal and we are unclear why this is the case.  For the MPI-only runs, some possibilities may be the MPI rank affinity/binding leading to each socket requiring memory from the other, or simply the memory bus becoming saturated (as PCG is a memory-bound algorithm). 

\begin{figure}[p]
\centering
\includegraphics[height=8in]{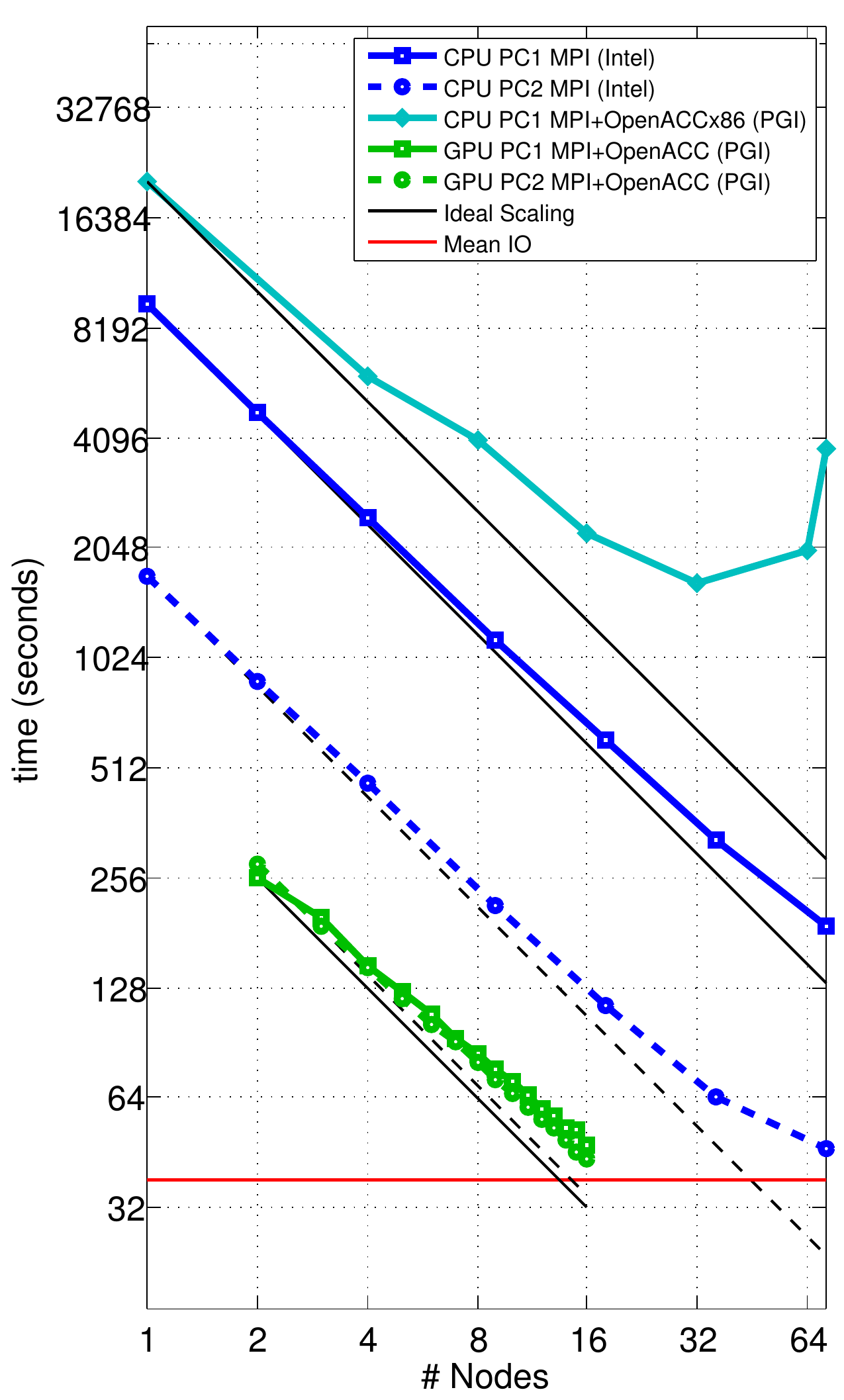}
\caption{CPU and GPU timing results of running POT3D on the test case with MPI-only and MPI+OpenACC for multiple nodes. Each CPU node has two 12-core Intel Haswell processors, while each GPU node has four NVIDIA Tesla P100 GPUs.\label{fig:time_cpugpu}} 
\end{figure}

In Fig.~\ref{fig:time_cpugpu}, we show performance results for POT3D using multiple CPU (and GPU) nodes.  Here, we see that the original MPI-only code scales very well using multiple CPU nodes up to about 64 nodes where it starts to lose efficiency. This loss of efficiency could be due to a combination of the number of iterations increasing (in the case of PC2) and load imbalance (made worse by the synchronizing nature of the inner products and polar-averaging boundary conditions).  

The PC2 preconditioner for the CPU runs is clearly much more efficient than PC1 (up to four times faster).  This should be compared to the reduction of iterations (nearly $10\times$) which shows the relative increase in computational expense of the PC2's application versus PC1.  We note that using the maximum number of CPU nodes available yields a computation time on the same order as the IO time (excluded from the timings) which is about as much performance as we can expect for POT3D in its current form.

For PC1, without any additional code changes, we can compile the MPI+OpenACC code to x86 multicore (in this case the OpenACC pragmas are treated by the compiler similar to OpenMP pragmas).  Although our focus is using OpenACC for GPU computing, we want to see what performance portability we can get using the same OpenACC source code on a different architecture.  Unfortunately, in this case, the code does not run efficiently, but for single-node runs, the results are not atypical to other codes we have tested comparing MPI-only with OpenMP-only.  This performance drop compared to the MPI-only code could be due to limits on the parallelization of small loops across single dimensions (such as boundary conditions and MPI send/receive buffers), and/or improper thread affinity and `first-touch' issues with data initialization.  For the multi-node MPI+OpenACCx86 runs, we set {\tt ACC\_NUM\_CORES} to 24 and the number of MPI ranks per node to 1.  The code shows decent scaling at first (taking into account that its computation time on one node is twice as slow as the MPI-only run).  However, as the number of nodes increases, the quality of the scaling decreases, and at high number of nodes, the performance stops improving and starts getting worse.  The timers in the code indicate that the vast majority of the time in these runs is being spent in the MPI calls (or waiting for synchronizations around them).  We are unclear why this is the case since the number of MPI ranks is far reduced compared to running with MPI-only. These MPI+OpenACCx86 results illustrate the difficulty in creating efficient hybrid MPI+multi-threaded CPU codes {--} often, a simple implementation does not yield good results.  Since our focus in this paper is using OpenACC for running our code on GPUs, a detailed MPI+OpenACCx86 performance analysis is beyond the current scope.

\subsection{GPU performance with MPI+OpenACC}
The GPU performance results using MPI+OpenACC on multiple nodes in Fig.~\ref{fig:time_cpugpu} show that the code scales well (especially considering that the system did not have NVlink).  The scaling of the code loses some efficiency when run on many nodes.  For PC2, the fact that we compute the PC2 formulation on the CPU before the start of the solver iterations may be affecting the scaling results.  We were not able to run the test case on less than 2 nodes due to the very large problem size (the GPUs did not have enough memory to contain the required arrays).  This limitation could be avoided by the use of unified memory, but doing so may hinder performance. 

Interestingly, we see that the PC1 and PC2 preconditioners have almost the same run time (sometimes the PC2 is even slower than PC1) which is indicative of the cuSparse LU-solver not being as relatively efficient for our matrix structure and/or storage format as in the CPU case.   Although the lack of increased performance of the PC2 preconditioner is not ideal, it does mean we can exclusively use the PC1 preconditioner.  This allows the code to be fully portable/single-sourced (as discussed in Sec.~\ref{sec:mpi2mpiopenacc}).

Comparing the GPU timings to the CPU results shows that, for this case, a single 4xP100 GPU node is roughly equal in performance to 20(4) 2x12-core Haswell nodes when using PC1(PC2).  Stating this in terms of single compute units, for PC1(PC2), one P100 GPU performs similarly to 10(2) 12-core Haswell CPUs.  Although these results are far less than the peak capabilities of the P100 GPUs, for a memory-bound code containing global collectives, they are not atypical.

We have used a very large problem size for these results because it is such large problems that drive the desire to use accelerated computing.  However, we do note that problem size can be very important for performance on GPUs since there must be enough work on the cards to get the best utilization.  We have previously tested POT3D on smaller problems and found the relative GPU vs CPU performance to be reduced somewhat (see Ref.~\cite{GTC17} for details).

\section{Performance Portability}
\label{sec:portable}
Performance portability is the idea of having one code base that can be run efficiently on multiple hardware architectures and types.  Various attempts at such portability have been proposed and implemented, but none has seemed to gain overwhelming acceptance and demonstrated true efficient portability.  

By its focus on expressing the parallelism in computational regions, OpenACC is designed to allow implementations to more easily target different hardware with minimal options in the code (such as conditional pragmas).  The PGI implementation of OpenACC targets numerous architectures including NVIDIA GPUs, multi-core x86 and OpenPower systems, and has plans to directly target Intel KNL nodes, as well as ARM and other systems.  Other implementations of OpenACC (such as GNU) also plan to target multiple devices.  

As the results in Sec.~\ref{sec:results} showed, the x86 multi-core implementation of OpenACC performed very poorly with our code/HPC system setup.  Perhaps modifying our run-time environment and/or adding initialization kernels could improve on the performance, but even if this is possible, it is unclear how optimizing the code for x86 would affect the GPU performance.

We did not have access to any OpenPower or KNL systems with the PGI compiler installed, so we were unable to test POT3D's performance portability on these architectures.

Although the OpenACC results for x86 multi-core were not efficient in this case, the OpenACC POT3D does exhibit performance portability as it runs very well using MPI-only on CPUs and with OpenACC on GPUs with the same source code (when using PC1).  
  
\section{Summary}
\label{sec:conclude}
We have demonstrated adding OpenACC accelerator support to a legacy FORTRAN MPI PCG solver code.  We can now analyze our original goals in light of the results:
\begin{enumerate}
\item To efficiently compute solutions in-house using a modest number of GPUs that would normally require the use of a multi-CPU HPC cluster
\end{enumerate}
\hangindent=0.75cm 
\hangafter=0
From our multi-node GPU results, we found that one P100 GPU was similar in performance to 5 24-core Haswell nodes for PC1 and 1 24-core Haswell node for PC2.  Since there are cases where PC1 is more efficient than PC2 (even on the CPU), a single 16-GPU workstation could perform as well as 80 Haswell nodes, which is more than the maximum number of nodes usable for a run on Comet.  Even for PC2 runs, having a 16-GPU server would be equivalent to 16 Haswell nodes, which is fast enough for computing medium-sized problems efficiently.

\begin{enumerate}
\setcounter{enumi}{1}
\item To compute large solutions more quickly that current HPC resources allow
\end{enumerate}
\hangindent=0.75cm 
\hangafter=0
On the NVIDIA PSG cluster, we were able to use up to 16 nodes (a total of 64 P100 GPUs).  The run using all 16 nodes was faster than that using the maximum 72 CPU nodes on Comet (for PC1, it was 4 times faster, for PC2 is was marginally faster).  Comet recently added new GPU-enabled racks with 4 P100 GPUs per node which will allow us to solve the solutions faster than we can on the CPU nodes.

\begin{enumerate}
\setcounter{enumi}{2}
\item To use less HPC allocation resources (i.e. do an equivalent solve on fewer GPU nodes than CPU nodes)
\end{enumerate}
\hangindent=0.75cm 
\hangafter=0
From the above analysis, if a 4xP100 GPU node had the same allocation cost as a CPU node, then the OpenACC version of POT3D would indeed save a significant amount of allocation time.  However, typically HPC systems charge more allocation for more advanced nodes (like the P100 GPU nodes) so the savings in allocation will vary depending on each HPC center's policies.
\linebreak

In addition to the goals above, it was very important that the resulting code be single-source (portable).  For PC1, this was achieved and the OpenACC enhanced code is now part of the main code base.  However, PC2 necessitated a break in portability due to the use of the cuSparse library.  The performance results in Sec.~\ref{sec:results} indicate that for the P100 GPUs, the performance of using PC1 and PC2 are roughly equivalent, in which case having OpenACC on only PC1 is acceptable.  This result is reminiscent of computations done in the late 1980s using Cray (and other) vector computers, where the value of incomplete factorization preconditioners versus the improved vector performance of diagonal scaling was discussed in the literature often (e.g. see Ref.~\cite{pini1990simple}).  With the advent of GPUs and other large-vector hardware (such as the Intel KNL processor with the AVX512 instruction set), renewed research into alternative vector-friendly preconditioners could be of great value. 

Overall, adding OpenACC to POT3D to add multi-GPU capabilities did not involve overly extensive development time, and for PC1 only required a 3\% change in the source code.  This, combined with the positive GPU performance results, encourages us to move forward in implementing OpenACC in our global MHD code MAS.   The lessons learned in the POT3D implementation will be of great assistance in that venture, and we hope they will also be of assistance to other researchers considering adding accelerated computing to their legacy code using OpenACC.
 
\section*{Acknowledgements}
Funding for this work was provided by NASA, NSF, and AFOSR.  Computer time on Comet was provided by the Extreme Science and Engineering Discovery Environment (XSEDE), which is supported by NSF grant number ACI-1548562.  We gratefully acknowledge NVIDIA Corporation for allowing us to utilize their PSG cluster for the P100 GPU runs.
 
\bibliographystyle{splncs03} 
\bibliography{POT3DGPU}

\begin{thebibliography}{10}
\providecommand{\url}[1]{\texttt{#1}}
\providecommand{\urlprefix}{URL }

\bibitem{arge2003improved}
Arge, C.N., Odstrcil, D., Pizzo, V.J., Mayer, L.R.: Improved method for
  specifying solar wind speed near the sun. In: AIP Conference Proceedings.
  vol. 679, pp. 190--193. AIP (2003)

\bibitem{DIACSR}
Bell, N., Garland, M.: Efficient sparse matrix-vector multiplication on {CUDA}.
  Tech. rep., Nvidia Technical Report NVR-2008-004, Nvidia Corporation (2008)

\bibitem{PC_iterative_survey}
Benzi, M.: Preconditioning techniques for large linear systems: a survey.
  Journal of computational Physics  182(2),  418--477 (2002)

\bibitem{ASTRONUM16}
Caplan, R.M., Mikić, Z., Linker, J.A., Lionello, R.: Advancing parabolic
  operators in thermodynamic {MHD} models: {E}xplicit super time-stepping
  versus implicit schemes with {K}rylov solvers. Journal of Physics: Conference
  Series  837(1),  012016 (2017)

\bibitem{GTC17}
Caplan, R., Linker, J., Mikic, Z.: Potential field solutions of the solar
  corona: Converting a {PCG} solver from {MPI} to {MPI}+{O}pen{ACC} (2017),
  \url{http://on-demand.gputechconf.com}, presented at the NVIDIA GPU
  Technology Conference

\bibitem{OpenACCBook2}
Chandrasekaran, S., Juckeland, G.: OpenACC for Programmers Concepts and
  Strategies. Addison-Wesley Professional (2017)

\bibitem{chapman2008using}
Chapman, B., Jost, G., Pas, R.v.d.: Using {O}pen{MP}: portable shared memory
  parallel programming. The MIT Press (2008)

\bibitem{ILU_breakdown}
Chow, E., Saad, Y.: Experimental study of {ILU} preconditioners for indefinite
  matrices. Journal of Computational and Applied Mathematics  86(2),  387--414
  (1997)

\bibitem{cook2012cuda}
Cook, S.: {CUDA} Programming: {A} Developer's Guide to Parallel Computing with
  {GPU}s. Morgan Kaufmann Publishers Inc. (2012)

\bibitem{downs2013probing}
Downs, C., Linker, J.A., Miki{\'c}, Z., Riley, P., Schrijver, C.J.,
  Saint-Hilaire, P.: Probing the solar magnetic field with a sun-grazing comet.
  Science  340(6137),  1196--1199 (2013)

\bibitem{farber2016parallel}
Farber, R.: Parallel Programming with OpenACC. Morgan Kaufmann, Elsevier (2016)

\bibitem{kaeli2015heterogeneous}
Kaeli, D.R., Mistry, P., Schaa, D., Zhang, D.P.: Heterogeneous Computing with
  {O}pen{CL} 2.0. Morgan Kaufmann Publishers Inc. (2015)

\bibitem{linker2016empirically}
Linker, J.A., Caplan, R.M., Downs, C., Lionello, R., Riley, P., Mikic, Z.,
  Henney, C.J., Arge, C.N., Kim, T., Pogorelov, N.: An empirically driven
  time-dependent model of the solar wind. In: Journal of Physics: Conference
  Series. vol. 719, pp. 12012--12023. IOP Publishing (2016)

\bibitem{MAS}
Lionello, R., Linker, J.A., Miki{\'c}, Z.: Multispectral emission of the sun
  during the first whole sun month: Magnetohydrodynamic simulations. The
  Astrophysical Journal  690(1),  902 (2009)

\bibitem{pini1990simple}
Pini, G., Gambolati, G.: Is a simple diagonal scaling the best preconditioner
  for conjugate gradients on supercomputers? Advances in Water Resources
  13(3),  147--153 (1990)

\bibitem{IterativeMethods_SAAD_Book}
Saad, Y.: Iterative methods for sparse linear systems. Siam (2003)

\bibitem{1970Natur226251S}
{Schatten}, K.H.: {Prediction of the Coronal Structure for the Solar Eclipse of
  March 7, 1970}. Nature  226,  251 (1970)

\bibitem{CSRopt}
Smith, B., Zhang, H.: Sparse triangular solves for ilu revisited: data layout
  crucial to better performance. International Journal of High Performance
  Computing Applications  25(4),  386--391 (2011)

\bibitem{titov20122010}
Titov, V., Mikic, Z., T{\"o}r{\"o}k, T., Linker, J., Panasenco, O.: 2010 august
  1-2 sympathetic eruptions. i. magnetic topology of the source-surface
  background field. The Astrophysical Journal  759(1), ~70 (2012)

\bibitem{toth2011obtaining}
T{\'o}th, G., Van~der Holst, B., Huang, Z.: Obtaining potential field solutions
  with spherical harmonics and finite differences. The Astrophysical Journal
  732(2),  102 (2011)

\bibitem{NUG_PLAY_1992}
Veldman, A., Rinzema, K.: Playing with nonuniform grids. Journal of engineering
  mathematics  26(1),  119--130 (1992)

\bibitem{wang1995solar}
Wang, Y.M., Sheeley~Jr, N.: Solar implications of ulysses interplanetary field
  measurements. The Astrophysical Journal Letters  447(2),  L143 (1995)

\end{thebibliography}
 
\end{document}